\documentclass[sigconf]{acmart}

\AtBeginDocument{%
  \providecommand\BibTeX{{%
    \normalfont B\kern-0.5em{\scshape i\kern-0.25em b}\kern-0.8em\TeX}}}

\copyrightyear{2023}
\acmYear{2023}
\setcopyright{acmlicensed}\acmConference[WWW '23]{Proceedings of the ACM Web Conference 2023}{May 1--5, 2023}{Austin, TX, USA}
\acmBooktitle{Proceedings of the ACM Web Conference 2023 (WWW '23), May 1--5, 2023, Austin, TX, USA}
\acmPrice{15.00}
\acmDOI{10.1145/3543507.3583526}
\acmISBN{978-1-4503-9416-1/23/04}

\usepackage{lipsum}
\usepackage{caption}
\usepackage{bm}
\usepackage{amsmath}
\usepackage{amsthm}
\usepackage{amsfonts}
\usepackage[toc,page]{appendix}

\usepackage{algorithm}
\usepackage{algorithmic}
\usepackage{booktabs}
\usepackage{xcolor}
\usepackage{subfigure}
\usepackage{multirow}
\usepackage{color}

\begin{document}

\title{Dual Intent Enhanced Graph Neural Network for Session-based New Item Recommendation}

\settopmatter{authorsperrow=4}

\author{Di Jin}
\affiliation{%
  \institution{College of Intelligence and Computing, Tianjin University}
  \country{China}}
\email{jindi@tju.edu.cn}

\author{Luzhi Wang}
\authornote{Corresponding author.}
\affiliation{%
  \institution{College of Intelligence and Computing, Tianjin University}
  \country{China}}
\email{wangluzhi@tju.edu.cn}

\author{Yizhen Zheng}
\affiliation{%
  \institution{Department of Data Science and AI, Faculty of IT, Monash University}
  \country{Australia}
}
\email{yizhen.zheng1@monash.edu}

\author{Guojie Song}
\affiliation{%
  \institution{School of Intelligence Science and Technology, Peking University}
  \country{China}
}
\email{gjsong@pku.edu.cn}

\author{Fei Jiang}
\affiliation{%
  \institution{Meituan}
  \country{China}
}
\email{f91.jiang@gmail.com}

\author{Xiang Li}
\affiliation{%
  \institution{Beijing}
  \country{China}
}
\email{leo.lx007@qq.com}

\author{Wei Lin}
\affiliation{%
\institution{Beijing}
  \country{China}
}
\email{lwsaviola@163.com}

\author{Shirui Pan}

\affiliation{%
  \institution{Griffith University}
  \country{Australia}
}
\email{s.pan@griffith.edu.au}

\renewcommand{\shortauthors}{Jin et al.}






\newcommand{\ourmethod}{\texttt{GCL-GP}}
\newcommand{\tabincell}[2]{\begin{tabular}{@{}#1@{}}#2\end{tabular}}

\begin{abstract}
    Recommender systems are essential to various fields, e.g., e-commerce, e-learning, and streaming media. At present, graph neural networks (GNNs) for session-based recommendations normally can only recommend items existing in users' historical sessions. As a result, these GNNs have difficulty recommending items that users have never interacted with (new items), which leads to a phenomenon of information cocoon. 
Therefore, it is necessary to recommend new items to users. As there is no interaction between new items and users, we cannot include new items when building session graphs for GNN session-based recommender systems. Thus, it is challenging to recommend new items for users when using GNN-based methods.
We regard this challenge as ``\textbf{G}NN \textbf{S}ession-based \textbf{N}ew \textbf{I}tem \textbf{R}ecommendation (GSNIR)''. 
To solve this problem, we propose a dual-intent enhanced graph neural network for it. 
Due to the fact that new items are not tied to historical sessions, the users' intent is difficult to predict. We design a dual-intent network to learn user intent from an attention mechanism and the distribution of historical data respectively, which can simulate users' decision-making process in interacting with a new item.
To solve the challenge that new items cannot be learned by GNNs, inspired by zero-shot learning (ZSL), we infer the new item representation in GNN space by using their attributes. 
By outputting new item probabilities, which contain recommendation scores of the corresponding items, the new items with higher scores are recommended to users. 
Experiments on two representative real-world datasets show the superiority of our proposed method. The case study from the real-world verifies interpretability benefits brought by the dual-intent module and the new item reasoning module.

\end{abstract}



\maketitle

\section{Introduction}
Recommender systems have a major impact on a variety of fields, including e-commerce \cite{ge2020understanding, kersbergen2021learnings}, e-business \cite{dias2008value}, and media streaming services \cite{wang2021clicks}, which can influence our purchasing decisions and even lifestyles. In recommender systems, the session-based recommendation is a common task \cite{wu2019session, wang2020global}. Based on the previous items a user has interacted with in a session (i.e., a sequence of items), session-based recommendations aim at predicting the next item a user will interact with. Early approaches are based on identifying frequent sequential patterns, which are used for next-item recommendations in e-commerce or music domains \cite{wang2021denoising,DBLP:journals/csur/BonninJ14}. Such methods are computationally expensive because they are parameter-sensitive algorithms \cite{DBLP:journals/umuai/LudewigJ18}. In recent years, some works have used Recurrent Neural Networks (RNNs) for a session-based recommendation. For example, Hidasi et al. \cite{DBLP:journals/corr/HidasiKBT15} use Gated Recurrent Units (GRUs) as a special form of RNNs to model sparse session data to predict the next user interactions in a session.
However, recommender systems based on RNNs have some limitations. For example, the number of interactions a user makes in a session is limited. It is difficult for RNNs to accurately estimate each user's preference from each session \cite{wu2019session}. Recently, graph neural networks (GNNs) are developing very fast and have been applied in many real-world applications~\cite{tan2023federated, liuyue_DCRN, zheng2022rethinking, zheng2022unifying}. In session-based recommendations, GNN-based approaches become increasingly popular as they can address the aforementioned issues and improve prediction results.

\begin{figure}
\centering
\includegraphics[scale=0.38]{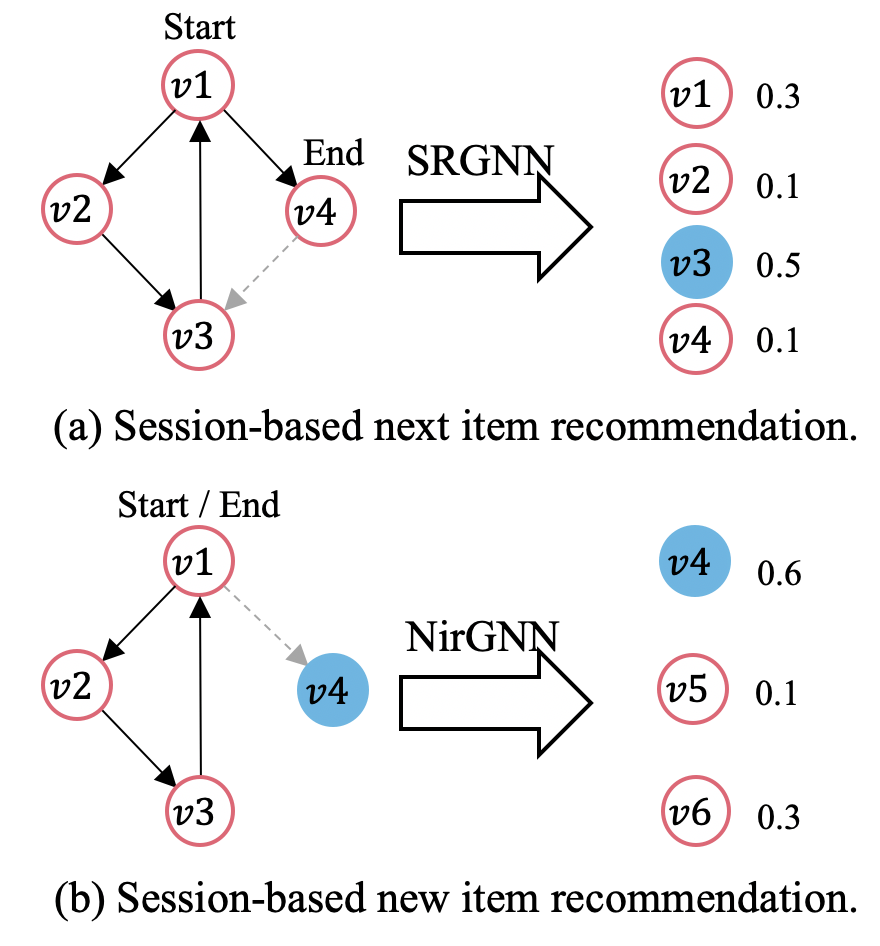}
\vspace{-1em}
\caption{Difference between session-based next item recommendation and GSNIR. The dashed line points to the ground truth items. In (a), the session is $v_1 \rightarrow v_2 \rightarrow v_3\rightarrow v_1 \rightarrow v_4$, and $v_3$ is the ground truth item. SR-GNN calculates the probability of each item that is liked by the user. In (b), $v_1,v_2,v_3$ are old items, $v_4,v_5, v_6$ are candidate new items, where $v_4$ is the ground truth new item. NirGNN calculates the selected probability by user for every candidate new items. 
}
\label{newrec} \vspace{-2em}
\end{figure}

Currently, there are many GNN-based methods for session-based recommendation, such as SR-GNN \cite{wu2019session}, GCE-GNN \cite{wang2020global} and GC-SAN \cite{xu2019graph}. These GNN-based models typically construct a user's historical sessions as a session graph. Then, a well-designed graph neural network learns the representations of users' intent and items from the historical session graph, and computes user-item matching scores. Finally, the recommender system calculates the probability that the user's preference for items based on the match score. However, these methods only recommend items that already exist in the session (old items), which can easily reduce the desire of users to interact with items and lead to information cocoons. Specifically, these methods take in the items in the user's historical session as input and output the probability of historical items to be recommended as the next recommended items. For example, in Fig. \ref{newrec} (a), the input of SR-GNN is $v_1, v_2, v_3, v_4$, and its output is the probability of $v_1, v_2, v_3, v_4$.
Therefore, users can be tired of the old items easily. 

To expand users' selection for session-based recommendations, we introduce a new challenge ``\textbf{G}NN \textbf{S}ession-based \textbf{N}ew \textbf{I}tem \textbf{R}ecommendation (GSNIR)''. In specific, GSNIR refers to predicting whether a user will interact with a new item in the future (e.g., purchase and click) based on his or her previous interaction. In Fig. \ref{newrec}, we present the comparison between the traditional GNN-based methods (e.g. SR-GNN) for session-based recommendation and our proposed new setting. Even though new item recommendation is very important to improve user experience, existing GNN-based models perform poorly in this setting. As we mentioned before, these GNN methods recommend old items, and they learn old item embeddings in session graphs. But for a new item that the user has not interacted with, it is not in the session, so it cannot use topology information to learn the embedding. Therefore, it is necessary to design a new GNN-based recommender system to solve the GSNIR.

The GSNIR meets many challenges: \textbf{\textit{Challenge 1. }}
How to simulate the decision-making process of users when they choose new items? When a user interacts with a new item, the user may tend to substitute the old item with its relevant attributes or look for substitutes in a large taxonomy. For example, if a user often buys coca-cola, when he/she buys a new product, he/she may choose pepsi as a substitution. Moreover, he/she may look for alternatives under the taxonomy subtree for beverages and choose coconut water. By learning user intent, we can filter new items that match the user's preference. However, human behavior is ever-changing, which makes it difficult to learn user intent accurately from the user's historical session. \textbf{\textit{Challenge 2.}} How can we obtain effective new item representations by GNNs? The new item recommendation problem poses a challenge to learning representation based on limited information. 
As there is no interaction between new items and users, it is difficult to include new items when building a session graph for GNN encoding.

To address the aforementioned problem, in this paper, we propose a novel GNN-based model, namely \textbf{N}ew \textbf{I}tem \textbf{R}ecommendation \textbf{G}raph \textbf{N}eural \textbf{N}etwork (NirGNN). This model can be divided into two components. The first component is to learn the user intent effectively. To do this, NirGNN proposes a dual-intent learning method.
NirGNN introduces the taxonomy tree of each item to determine the general direction of the user intent. By using both the attention mechanism and the $\beta$ distribution on taxonomy data, the method predicts whether the user prefers the taxonomy for the interaction guide or the item itself.
The second component is to obtain new item representation. Inspired by zero-shot learning (ZSL), we design a reasoning method to map the independent new items into the GNN space by using item attributes, which enables new-item recommendation. To evaluate the effectiveness of NirGNN, we provide two new real-world datasets for the new item recommendation setting and conduct experiments on both datasets. Experimental results demonstrate the effectiveness of our method. Besides, we provide a case study in a commercial recommendation environment, which demonstrates the interpretability and practicability of our model. 
In summary, the contributions of this paper can be summarised as follows: 
\vspace{-0.5em}
\begin{itemize} 
    \item To the best of our knowledge, we are the first to propose the GNN session-based new item recommendation for the session-based recommendation scenario.
    \item We propose a dual-intent enhanced graph neural network to tackle GSNIR. The dual-intent strategy learns user intent from users' attention and data distribution respectively, which provides a different perspective on user preference.
    \item 
    We introduce zero-shot learning into a session-based recommender system for the first time, which solves the problem that new items cannot be encoded by GNN-based methods due to the lack of interaction with users.
    \item Extensive experiments conducted on session-based recommendation datasets show the superiority of NirGNN. Moreover, we present a real-world case study to demonstrate our method's interpretability and effectiveness.
\end{itemize}

\section{Related Study}
\textbf{GNNs for session-based recommendation.} Many types of session-based recommendation methods have been proposed, e.g., Markov chains \cite{zhanghe,DBLP:journals/tkde/JinYJPHWYZ23,yu2019multi}, and Recurrent Neural Networks (RNNs) \cite{liu2021case4sr,zhou2020cnn}. In recent years, to improve session-based recommendation, GNNs \cite{wang2023contrastive,DBLP:conf/ijcai/JinWZLJLP22, DBLP:conf/www/JinHL021, DBLP:conf/icdm/YuJLHWT021} have been incorporated into session-based recommender systems to model complex relations between adjacent items. For example, as the first work introduces GNNs into the session-based recommendation, SR-GNN \cite{wu2019session} transforms each session into a session graph and uses a GNN to capture the complex relations of items in the graph. 
GC-SAN \cite{xu2019graph} extends SR-GNN by incorporating multi-layer self-attention with residual connection to model the long-term preference.
GCE-GNN \cite{wang2020global} combines the global session graph with the local session graph and position embedding to better represent the feature of items. 
COTREC \cite{xia2021self} utilizes contrastive learning and develops a self-supervised learning with co-training method for session-based recommendations.

\noindent
\textbf{Zero-shot learning.} ZSL simulates the way of human reasoning to recognize new things that have never been seen before in CV \cite{chen2021free,han2021contrastive,naeem2021learning}, NLP \cite{cheng2021data,liu2021task,geng2021ontozsl}, and GNN \cite{mancini2022learning,gao2020ci,liu2020attribute}. For example, through the shape information of horses, and stripe information of tigers, the ZSL model is capable of identifying zebra images. 
Lampert \textit{et al.} \cite{lampert2009learning} encodes attribute representations of images and then uses these attribute embeddings to do classification inference based on human-specified attribute descriptions of the test images. ABS-Net \cite{lu2018attribute} generates pseudo feature representations for unseen images. With the pseudo feature as a data augmentation approach, ABS-Net can tackle both zero-shot learning and conventional supervised learning.

\vspace{0.5em}
\section{Problem Statement}

\subsection{Preliminary}
A user historical session $S=\{v_1,v_2,v_3,...,v_n\}$ consists of items that the user has interacted with (old items), each item $v_i$ in $S$ is sorted by timestamp, which rigidly indicates the order of items occurring in the $S$. The historical session can be modeled as a directed session graph $G=(V, E)$ where $|V|=n$, and an edge $e(v_i,v_j)\in E$ represents that the user interacts vertex $v_j$ after interacting vertex $v_i$. Given a set of new items $C=\{c_1,c_2,...,c_m\}$ that have never been interacted with, the GSNIR aims to recommend new items to users. Specifically, for any session graph $G$, the output of the session-based new item recommendation model is a ranked new item list with the predicted likelihood of each new item. The $top-k$ new items with the highest probability in $C$ will be recommended to the user. Each item $v_i$ contains a taxonomy tree, including three different levels, namely $t_{i1}, t_{i2}, t_{i3}$, representing different granularity levels from large to small. For example, in a food recommender system, the three taxonomy levels of a brand of coconut water are `Beverages', `Bottled Beverages', and `Coconut water'. The taxonomy information provides external information, which makes GNNs limit-free to ontology information of items in learning. In addition, each item contains a set of attribute information $A=\{a_{i1},...,a_{ik}\}$. For example, the attribute of a coconut water $v_i$ may include its brand and price attributes. Attribute information is beneficial for us to infer new item embeddings. The embedding of the three-level taxonomy tree of item $v_i$ can be regarded as $t_{i1},t_{i2},t_{i3}$, and the global taxonomy embedding of $v_i$ is $t_i = W(t_{i1},t_{i2},t_{i3})$, where $W \in \mathbb{R}^{3d\times d}$ is a Multi-layer Perceptron (MLP) to compress $t_{i1},t_{i2},t_{i3}$ into the latent space $\mathbb{R}^{d}$. For attributes of items in the session and the attributes of the new items, we use the word2vec model to process them. The attribute embedding of old items and new items are $atr$ and $atr^*$ respectively.

\subsection{Construct Session Graph}
\begin{figure}\vspace{-0.5em}
\centering
\includegraphics[scale=0.43]{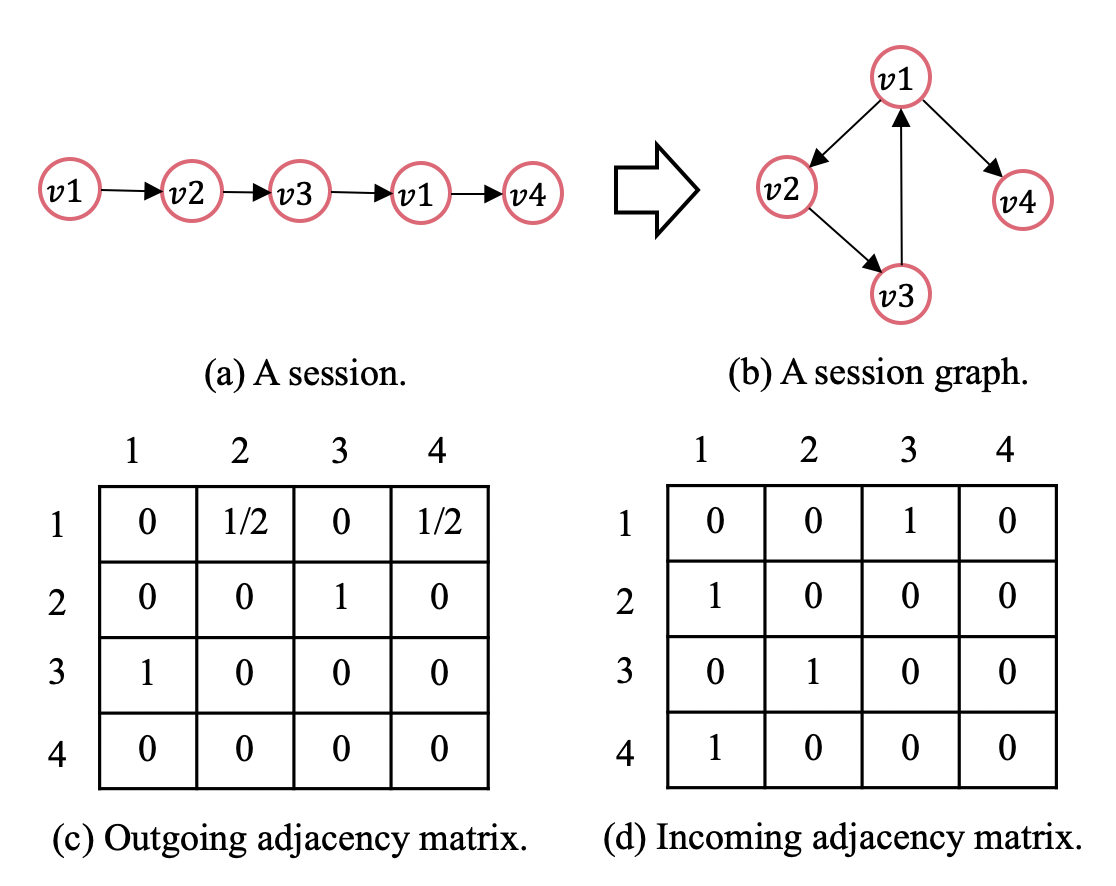}
\caption{A constructed session graph and the adjacency matrices. 
}
\label{segraph}\vspace{-1em}
\end{figure}

We follow the method proposed by SR-GNN to construct the session graph. The items in the session sequence are arranged in order. Each item is denoted as a vertex of the session graph, and each item and its next interacted item form an edge. Edges are divided into in-degree and out-degree.
For example, in $v_1 \rightarrow v_2$, $v_1$ is a start node and $v_2$ is an end node of the edge. This edge is the outgoing edge of $v_1$ and the incoming edge of $v_2$. Therefore, two adjacency matrices are needed to represent the session graph.
Each edge is assigned a normalized weight. The weight is calculated as the occurrence of the edge divided by the indegree/outdegree of that edge’s start node. For example, given a session $v_1 \rightarrow v_2 \rightarrow v_3 \rightarrow v_1 \rightarrow v_4$, Fig. \ref{segraph} (a) (b) shows the details of session graph constructing. In the outgoing adjacency matrix, the out-degree of $v_1$ is $2$, and $v_1\rightarrow v_2$ occurs once, then the weight of $v_1\rightarrow v_2$ is $\frac{1}{2}$. An example of an adjacency matrix is shown in Fig. \ref{segraph}. Fig. \ref{segraph} (c) (d) is the outgoing adjacency matrix and incoming adjacency matrix of (b), respectively. The outgoing adjacency matrix and incoming adjacency matrix together construct adjacency matrices for the session graph.

\begin{figure*}[htbp]
	\centering
	\includegraphics[width=0.9\textwidth]{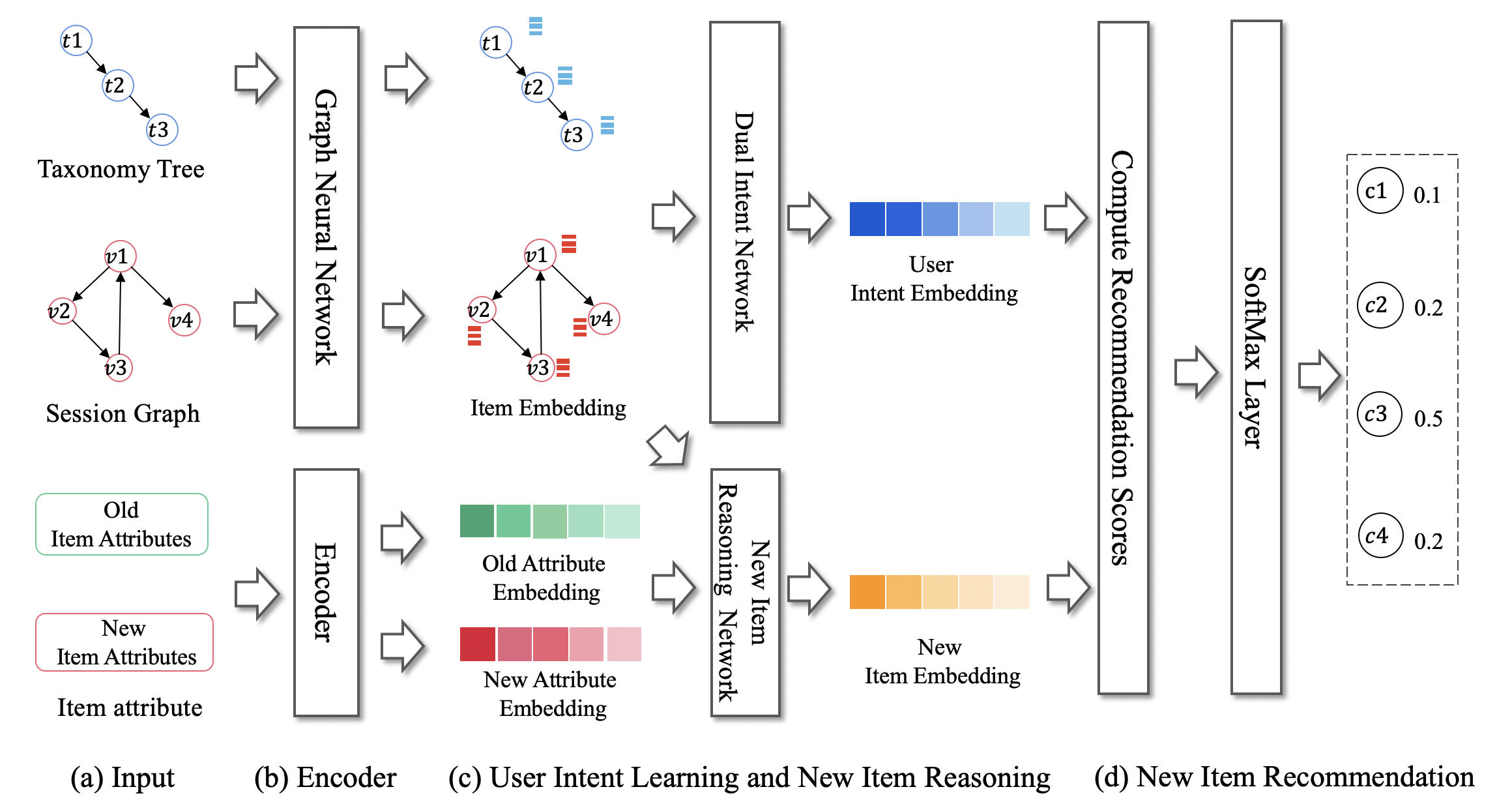} 
	\caption{Overview of NirGNN. In the first part of inputs (a), the taxonomy tree of each item and the session graphs are fed into a GNN encoder (b) to obtain the representation, and then (c) the dual-intents of the user are learned from the perspective of attention and data distribution. In the second part of inputs (a), attributes of items in the session graph and new item attributes are input into a word embedding encoder (b). Then, the semantic information of every item can be learned. Through a new item reasoning network (c), the learned semantic information is used to infer new item embedding. Finally, the new item embedding and user intent are used to calculate the score of each new item (d), and the new item with the highest score is regarded as the final recommendation. We introduce the four modules separately in the following sections.}
	\label{flowchart}
\end{figure*} 

\subsection{Problem Setting}
The general session-based GNN recommender system is usually the next item recommendation. That is, the input is a session containing items, and the output is the predicted score for each item in the session.
However, the input of GSNIR is the historical session, and the output is the predicted score of the candidate new items, which have not appeared in the historical records.

Since we cannot know what new items the user will interact with, we take the last item of the historical session as the new item (ground truth) of recommendation and mask it from the session. 
To ensure that the new item appears for the first time to the user, we delete it from the historical session and reconnect the historical session in the order in which the item was interacted with by the user. 

\section{Method}
\label{method}

This section details our proposed \textbf{N}ew \textbf{I}tem \textbf{R}ecommendation \textbf{G}raph \textbf{N}eural \textbf{N}etwork (NirGNN) model. 
Fig. \ref{flowchart} shows the overview of NirGNN.


\vspace{-0.5em}

\subsection{Item Embedding}
Since both the session graph and the taxonomy tree are directed graphs, we use the session graphs constructing approach of SR-GNN \cite{wu2019session} to process the adjacency matrix of directed graphs. Then, we use a gated graph session neural network (GGNN) as the core encoder to learn the representation of each item and its taxonomy tree. The learning process of GGNN can be summarised as follows \cite{wu2019session}:
\begin{equation} \label{XXX}
	\begin{split}
		a^{(t)}_{i} & = A_{i:} \left[v^{(t-1)}_1, \dots ,v^{(t-1)}_{n}\right]^\top H + b,\\
    	z^{(t)}_{i} & = \sigma\left(W_za^{(t)}_{i}+U_zv^{(t-1)}_{i}\right),\\
    	r^{(t)}_{i} & = \sigma\left(W_ra^{(t)}_{i}+U_rv^{(t-1)}_{i}\right),\\
    	\widetilde{v^{(t)}_{i}} & = \tanh\left(W_o a^{(t)}_{i}+U_o \left(r^{(t)}_{i} \odot v^{(t-1)}_{i}\right)\right),\\
    	v^{(t)}_{i} & = \left(1-z^{(t)}_{i} \right) \odot v^{(t-1)}_{i} + z^{(t)}_{i} \odot \widetilde{v^{(t)}_{i}},
	\end{split} 
\end{equation}
where \(\left[v^{(t-1)}_1, \dots , v^{(t-1)}_{n}\right]\) is the list of item vectors in session $s$, \(t\) is the training step, \(A_{i:} \in \mathbb{R}^{1 \times 2n}\) is the \(i\)-th row in matrix, \(H \in \mathbb{R}^{d \times 2d}\) and \(b \in \mathbb{R}^d\) are weight and bias parameter respectively, \(z_{i} \in \mathbb{R}^{d \times d}\) and \(r_{i} \in \mathbb{R}^{d \times d}\) are the reset and update gates respectively, \(\sigma(\cdot)\) is the sigmoid function, and \(\odot\) denotes element-wise multiplication. For each session graph $G$, the GGNN model propagates information between neighboring items. 

\subsection{Dual-Intent Nets}
We learn the intent of the user from the perspective of an attention mechanism and data distribution in the session graph respectively. The following is a detailed introduction.

\subsubsection{User $\alpha$ Intent Net}
We first learn user intent from a soft-attention mechanism perspective. Since most items interacted by the user recently are more likely to reflect the short-term preferences of the user than the items interacted at an earlier time, we extract the information of the last-interacted item $v_n$ and add it to each item. In addition, the taxonomy tree information also provides a macro guide for user intent. For a session graph, the intent of the user $I_{\alpha}$ can be learned as:
\begin{equation}
    I_{\alpha} = \sum_{i=1}^{n}\sigma((v_i \oplus t_i)*W_1+(v_n \oplus t_n)*W_2)(v_i \oplus t_i),
\end{equation}
where $\sigma(\cdot)$ is a sigmoid function, $\oplus$ is a concatenation function, $t_n$ is the global taxonomy embedding of $v_n$, $W_1$ and $W_2$ are the two intent filtering matrix. By applying the soft-attention mechanism to the session graph, we obtain the $\alpha$ intent of the user.


\subsubsection{User $\beta$ Intent Net}
In a recommender system, the historical session of the user, and the taxonomy tree/attributes of every item may be incomplete. To effectively mine the user's intent, only using the user $\alpha$ intent net is inadequate. The Bayesian inference method is often used to make the information complete by reasoning on new information. However, Bayesian inference is a very complicated process that requires complex calculations. It is possible to reduce the computational difficulty of the Bayesian inference if we use the $\beta$ distribution. 
The $\beta$ distribution is conjugated before the binomial likelihood. We can use the $\beta$ distribution in place of the prior distribution, since the posterior distribution also follows the $\beta$ distribution, which reduces the cost of computation. Moreover, the $\beta$ distribution is a superposition of multiple binomial distributions, which is also reflected in the conjugation of the binomial distribution. For example, the daily behavior of users (interacting or not interacting) is a binomial distribution, and many days are superimposed together to form a $\beta$ distribution. Thus, NirGNN calculates the posterior distribution by item embedding $v_i$ and taxonomy embedding $t_i$ without multiplying the likelihood by the prior assignment for user intent. 
We consider the user's preference from the perspectives of $v_i$ and $t_i$, and learn whether the user prefers the item itself or its taxonomy tree. $v_i$ and $t_i$ constitute a set of joint distribution probabilities. The $\beta$ distribution represents the probability that the user likes the joint distribution probability of $v_i$ and $t_i$, that is, the probability that the user likes $v_i$ and $t_i$ in a certain distribution state. For each vertex $v_i$, user intent distribution $b_i$ is:


\begin{equation}
\begin{split}
    b_i &= \mathcal{S}(\delta(v_i,t_i) x^{\phi(\varrho(v_i))-1}(1-x)^{\phi(\varrho(t_i)) -1}), \\
    \vspace{0.5em}
    \delta &= \frac{\Gamma(\phi(\varrho(v_i)) + \phi(\varrho(t_i)))}{\Gamma(\phi(\varrho(v_i)))\Gamma(\phi(\varrho(t_i)))},
\end{split}
    \label{7}
\end{equation}
where $\mathcal{S}(\cdot)$ is the sample function, $x$ is a variable, $\mathrm{\Gamma(\cdot)}$ is the Gamma distribution, $\varrho(\cdot)$ is a Softplus activation function, $\phi(x)=log(1+e^x)$ is a normalization function used to adjust the distribution interval of the data. Specifically, we use the activation function Softplus to process the embeddings of $v_i$ and $t_i$ and normalize them to the range of $(0, 1)$. We mainly consider whether the user is more interested in the representation of $v_i$ or the representation of $t_i$. The joint probability of the representation of $v_i$ and the representation of $t_i$ constitutes a set of probability density functions. The $\beta$ distribution shows the probability distribution that users are interested in $v_i$ or $t_i$ under different parameters. We find the closest probability to the user's true intent with sampling on $\beta$ distribution. Therefore, from the perspective of data distribution, $\beta$ distribution reflects the joint probability distribution of item and taxonomy tree. From the perspective of multiple cumulative distributions, it also reflects the intent of the user.

In a session graph, the recently interacted items are often more reflective of the user's recent interest than the items that interacted a long time ago. Therefore, we add a soft-attention mechanism with $\beta$ distribution to increase the priority of recently interacted item $v_i$. 
\begin{equation}
v_i' = b_i*(v_i\oplus t_i)+ b_n*(v_n \oplus t_n),
\label{8}
\end{equation}
where $v_i'$ is the embedding of $v_i$ that is prioritized.
However, the above Eq.\ref{8} will weaken the probability of $v_i$, so we readjust the probability distribution of each vertex so that they re-obey the $\beta$ distribution, which is equivalent to obtaining the attention that obeys the $\beta$ distribution. The attention modified $\beta$ distribution of each item $v_i$ is:


\begin{equation} \label{XXX}
        \beta_i =  \frac{v_i'-\frac{1}{n}\sum_{i=1}^{n}(b_{i}*(v_{i}\oplus t_{i})) }{
        \sqrt{\frac{\sum_{i=1}^n(b_i-\mathrm{avg}(b))^2}{n}}}W_3,
\end{equation}
where $\mathrm{avg(\cdot)}$ is the average function, and $W_3$ is a MLP to dimensionality reduction. After we get the modified $\beta$ distribution, we can get the user intent expressed by items in the session graph. The $\beta$ intent of the user can be defined as: 
\begin{equation}
    I_{\beta} = \sum_{i=1}^{n}(v_i\oplus t_{i})*\beta_i.
\end{equation}
By combining the intent from the two perspectives of $\alpha$ intent and $\beta$ distribution intent, we design a weighted intent $I$: 
\begin{equation}
    I = \lambda   I_{\alpha}  + (1-\lambda)  I_{\beta},
\end{equation}
where $\lambda \in [0,1]$ is a parameter that controls the weight of each intent.

\subsection{New Item Reasoning Network}

\begin{figure}\vspace{-0.2em}
\centering
\includegraphics[scale=0.4]{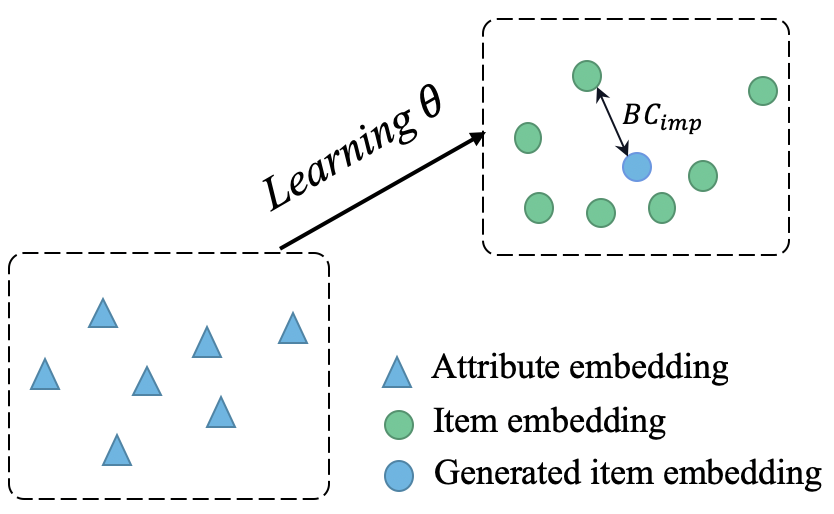}
\caption{The processing of new item reasoning. The left is the attribute space, and the right is the embedding space.
}
\label{zero}\vspace{-0.5em}
\end{figure}
Since the new items are time-independent, we cannot directly use GNN to embed the new item, which makes it impossible to calculate the similarity of user intent to the new item. To this end, we try to use the semantic information of the new item attribute to infer the embedding of the new item. Inspired by zero-shot learning, we can establish the spatial transformation relationship between item semantic attributes and item embedding. Specifically, given the attribute semantics $atr_i$ of old items $v_i$, we learn a transformation function $\theta$ to map the semantic attribute information to the item embedding space. Specifically, the generated item embedding can be calculated as $v^*_i= \theta(atr_i)$. We implement $\theta$ as a MLP layer. In order to ensure that the generated item embedding is in the same space as the original embedding, we try to reduce the gap between original embedding $v_{i}$ and generated embedding $v^*_i$. We measure the gap between them with an improved Bhattacharyya distance, which gives an upper bound on the minimum error rate for the separability of the two discrete probability distributions. The improved Bhattacharyya distance $BC_{imp}$ is:
\begin{align}
\begin{split}
BC_{imp}(v_i,v_i^*)= \left \{
\begin{array}{ll}
    -log(\sqrt{v_{i} * v_i^*}) & , \sqrt{v_{i} *  v_i^*} \neq \varnothing.\\
    0                               & , otherwise.
\end{array}
\right.
\end{split}
\end{align}
Fig. \ref{zero} shows the process of reasoning.
To improve the transformation function $\theta$ from semantics attribute space to item embedding space, we design a loss function $L_{zero}$ to optimize the difference between the original item embedding and the generated item embedding from the item semantics attributes.
\begin{equation}
   L_{zero}(v_i,v_i^*) = BC_{imp}(v_i, v_i^*).
\end{equation}
Then we use this transformation function to reason the new item embedding by the attribute information of the new item. The embedding of the new item $c_i$ can be inferred by the spatial transform function $c_i = \theta(atr_i^*)$, where $atr_i^*$ is the attribute embedding of new item $c_i$. 

\subsection{Prediction}
Now, we obtain the new item embedding $c_i$ and user intent from the above process. The recommendation score for each new item can be calculated by multiplying new item embeddings $c_i$ by user intent $I$, which can be defined as:
\begin{equation}
   \hat{Z_i} = \mathrm{softmax}(I^\top*c_i).
\end{equation}
The final loss consists of the CrossEntropy Loss $L_{ce}$, which is used to calculate the difference between the new item score and the ground truth, and $L_{zero}$, which is used to optimize the inference ability of the spatial transfer function:
\begin{equation}
   L = \gamma L_{ce}(Z,\hat{Z})+ (1-\gamma)(\sum_{i=1}^n L_{zero}(v_i, \theta(atr_i))),
\end{equation}
where $Z$ is the one-hot encoding vector of the ground truth item, $\hat{Z}$ denotes the recommendation scores for overall new items, and $\gamma \in [0,1]$ is the weight to control the importance of losses. 

\begin{table}[htbp]\vspace{-0.5em}
\centering
\caption{Dataset Description}
\resizebox{0.48\textwidth}{!}{
\begin{tabular}{@{}ccccc@{}}
\toprule
Datasets  & Items & Train sessions & Test sessions & Average length \\ \midrule
Amazon G\&GF  & 18889 & 51958          & 37813         & 8.394          \\
Yelpsmall & 14726 & 19035          & 2311          & 4.666          \\
\bottomrule
\end{tabular}
}
\label{dataset}\vspace{-1em}
\end{table}

\section{Experiment}

\begin{table*}
\centering
  \caption{Experiments on Amazon G\&GF and Yelpsmall dataset.}
  \label{sample-table}
  \resizebox{1\textwidth}{!}{
  \begin{tabular}{lllllllll}
    \toprule
    \multirow{2}{*}{Methods} & \multicolumn{4}{c}{Amazon G\&GF} & \multicolumn{4}{c}{Yelpsmall}       \\
    \cmidrule(r){2-9}
        & P@20             & MRR@20            & P@10              & MRR@10 & P@20             & MRR@20            & P@10              & MRR@10\\
    \midrule
    SR-GNN  & 1.438$\pm$0.134  & 0.336$\pm$0.025   & 0.086$\pm$0.078   & 0.297$\pm$0.021    & 0.764$\pm$0.534  & 0.192$\pm$0.118  & 0.476$\pm$0.346  & 0.180$\pm$0.111  \\
    SR-GNN-ATT    &0.678$\pm$0.202           &0.157$\pm$0.038            &0.375$\pm$0.144           & 0.136$\pm$0.040      & 1.631$\pm$0.353            &0.410$\pm$0.089             &0.989$\pm$0.183            &0.366$\pm$0.078 \\
    GC-SAN       &2.028$\pm$0.108           &0.580$\pm$0.093           &1.288$\pm$0.059           & 0.529$\pm$0.090     &0.793$\pm$0.115           &0.174$\pm$0.094           &0.418$\pm$0.144           & 0.165$\pm$0.103         \\   
    GCE-GNN & 1.650$\pm$0.291  &0.478$\pm$0.082  & 1.053$\pm$0.200  & 0.441$\pm$0.076 &1.702$\pm$0.462 & 0.789$\pm$0.224 &\textbf{1.428$\pm$0.563}  &\textbf{0.776$\pm$0.229} \\
    COTREC   & 1.681$\pm$0.837 & 0.434$\pm$0.229 & 1.078$\pm$0.583 & 0.393$\pm$0.213 & 0.833$\pm$0.417  & 0.256$\pm$0.199  & 1.021$\pm$0.583 & 0.269$\pm$0.222 \\     \midrule
    NirGNN(Ours)     & \textbf{2.420$\pm$0.039}  & \textbf{0.599$\pm$0.017}   & \textbf{1.578$\pm$0.056}   & \textbf{0.543$\pm$0.018}   & \textbf{1.817$\pm$0.087}  & \textbf{1.092$\pm$0.017}  & 0.779$\pm$0.043  & 0.299$\pm$0.049  \\  \bottomrule

  \end{tabular}
  \label{overallexp}
  }
\end{table*}

We run NirGNN on an NVIDIA GeForce RTX 3090 24G GPU for experiments. In the following section, we will introduce the description of the dataset, and present the experiment results of NirGNN for the new item recommendation, ablation study, and parameter analysis on these datasets.

\subsection{Dataset Description}
\subsubsection{Datasets.} The session-based user-oriented new item recommendation problem is a brand new problem and there are no public datasets suitable for it yet. Thus, we construct two real-world datasets to evaluate our model, including Amazon Grocery and Gourmet Food (Amazon G\&GF) dataset and Yelpsmall dataset. 
\begin{itemize}
    \item \textbf{Amazon G\&GF}. The Amazon G\&GF dataset is the subset of the Amazon dataset released in 2018\footnote{http://jmcauley.ucsd.edu/data/amazon/}. Item attributes are obtained from item data, including brands and prices.  The Amazon dataset describes the merchant with taxonomy, which uses a hierarchical taxonomy relationship, such as [`Sports \& Outdoors', `Other Sports', `Dance']. These taxonomy relationships can be obtained directly.
    \item \textbf{Yelpsmall}. The Yelpsmall dataset is from the 2021 release version of Yelp stored on the Kaggle\footnote{https://www.kaggle.com/datasets/yelp-dataset/yelp-dataset?datasetId=10100}. The Yelpsmall is created by randomly selecting 6000 users from the Yelp dataset with the review time in 2016. Based on the comment data of the two datasets, we obtained the session of user transactions and merchants sorted by time. The taxonomy in the Yelp dataset does not contain a hierarchical relationship. We use the following steps to generate the corresponding taxonomy tree. First, we pre-train the GoogleNews-vectors-negative300 by word2vec \cite{mikolov2013efficient} to obtain the word vectors of each taxonomy and then use K-Means++\cite{arthur2006k} to cluster word vectors. The same taxonomy's word vectors are averaged before being clustered once more. We first clustered $100$ taxonomy and then added the vectors of each taxonomy. These $100$-word vectors are clustered into $50$ taxonomy and then clustered into $10$ taxonomy, thus the three-level taxonomy tree is obtained. A three-level taxonomy tree is created after three clustering iterations. 
\end{itemize}
Based on the comment data of the two datasets, we obtained the session of user transactions and merchants sorted by time. In a session, a new item is defined as an item that has not appeared in the session. Therefore, we take the last item visited by the user according to timestamp as the ground truth for the new item recommendation, and delete it from the session to ensure that it is unique and new to the user. Following the data division method of SR-GNN, the session with a timestamp before $7$ days is divided into a training set, and the session after $7$ days is the test set. Table \ref{dataset} shows the details.
\subsubsection{Baselines.}
To evaluate the performance of the proposed method, we compare NirGNN with the following state-of-the-art baselines. For baselines, we use new items to calculate recommendation scores.
\begin{itemize}
\item \textbf{SR-GNN} \cite{wu2019session} is the first GNN-based model for the session-based recommendation. It uses gated GNN to encode items in each session and employs the attention mechanism to encode the global session preference. 
\item \textbf{SR-GNN-ATT} is SR-GNN with an attention mechanism.
\item \textbf{GC-SAN} \cite{xu2019graph} uses a gated GNN to encode items in each session and uses multi-layer self-attention with residual connection to encode the global session preference.
\item \textbf{GCE-GNN} \cite{wang2020global} uses a global session graph and a local session graph with position embedding.
\item \textbf{COTREC} \cite{xia2021self} is a self-supervised graph co-training framework. COTREC reports that the effect of COTREC is higher than that of the supervised SOTA session-based GNN model.
\end{itemize}

\subsubsection{Metrics.}
The following metrics are used to evaluate the comparative methods. $P@k$ (Precision@$k$) is widely used as a measure of prediction accuracy. It predicts the correct correlation result as a proportion of all returned results. $k$ represents the precision of the first $k$ items in the list. $MRR@k$ (Mean Reciprocal Rank) is the reciprocal rating of the re-averaged correctly recommended item. A large $P$ value or a large $MRR$ value indicates that the correct recommendation is at the top of the ranking list. In our experiments, we take the $k=\{10, 20\}$ as the evaluation index respectively. We use percentages to represent these data.

\begin{table}[htb]
\centering
\vspace{-0.5em}
\small
\footnotesize
\caption{Ablation study on Amazon G\&GF.}
\resizebox{0.48 \textwidth}{!}{
\begin{tabular}{lllll}
\toprule
Methods                &P@20  &MRR@20 &P@10  &MRR@10 \\ \hline
NirGNN & \textbf{2.420$\pm$0.039}  & \textbf{0.599$\pm$0.017}   & 1.578$\pm$0.056   & \textbf{0.543$\pm$0.018}   \\ 
NirGNN w/o $\alpha$ intent &2.410$\pm$0.030  &0.596$\pm$0.021  &1.508$\pm$0.048  &0.538$\pm$0.014 \\
NirGNN w/o $\beta$ intent  &2.375$\pm$0.019 &0.589$\pm$0.018 &1.512$\pm$0.008 &0.538$\pm$0.009  \\
NirGNN w/o $L_{zero}$    &2.237$\pm$0.119  &0.598$\pm$0.009  &\textbf{1.599$\pm$0.005} &0.541$\pm$0.010  \\ 
\bottomrule
\end{tabular}
}
\label{abs}\vspace{-1em}
\end{table}
\subsection{Evaluation Results}
\subsubsection{Overall Performance.} To demonstrate the overall performance of the proposed model, we compare it with five state-of-the-art session-based GNN recommendation methods. Here, we show new results on our proposed NirGNN framework along with baseline methods for different metrics. We show experimental results on Amazon G\&GF and Yelpsmall datasets in Table \ref{overallexp}, where the optimal solution for each metric is highlighted in bold and ‘$\pm$’ indicates a numerical range. We observe that NirGNN achieves significant performance gains under different metrics compared to the baseline. In particular, on the Amazon G\&GF dataset, in the case of $k=20$, NirGNN can improve by up to $19.33\%$ on $P$ metrics and improve $3.28\%$ on $MRR$ metrics which is compared to the second-best baseline. In the case of $k=10$, NirGNN can improve by up to $22.52\%$ on $P$ metrics and $2.65$ on $MRR$ metrics. On the Yelpsmall dataset, in the case of $k=20$, NirGNN can be improved by up to $38.40\%$ on $MRR$ metrics, and improved by up to $6.75\%$ on $P$ metrics compared to the second-best baseline. These experimental results demonstrate that the NirGNN model can effectively improve the learning performance for the new item recommendation.

\begin{figure}[ht]\vspace{-1em}
	\centering
	
	\subfigure[{$\lambda$.}]{\includegraphics[width=0.49\columnwidth]{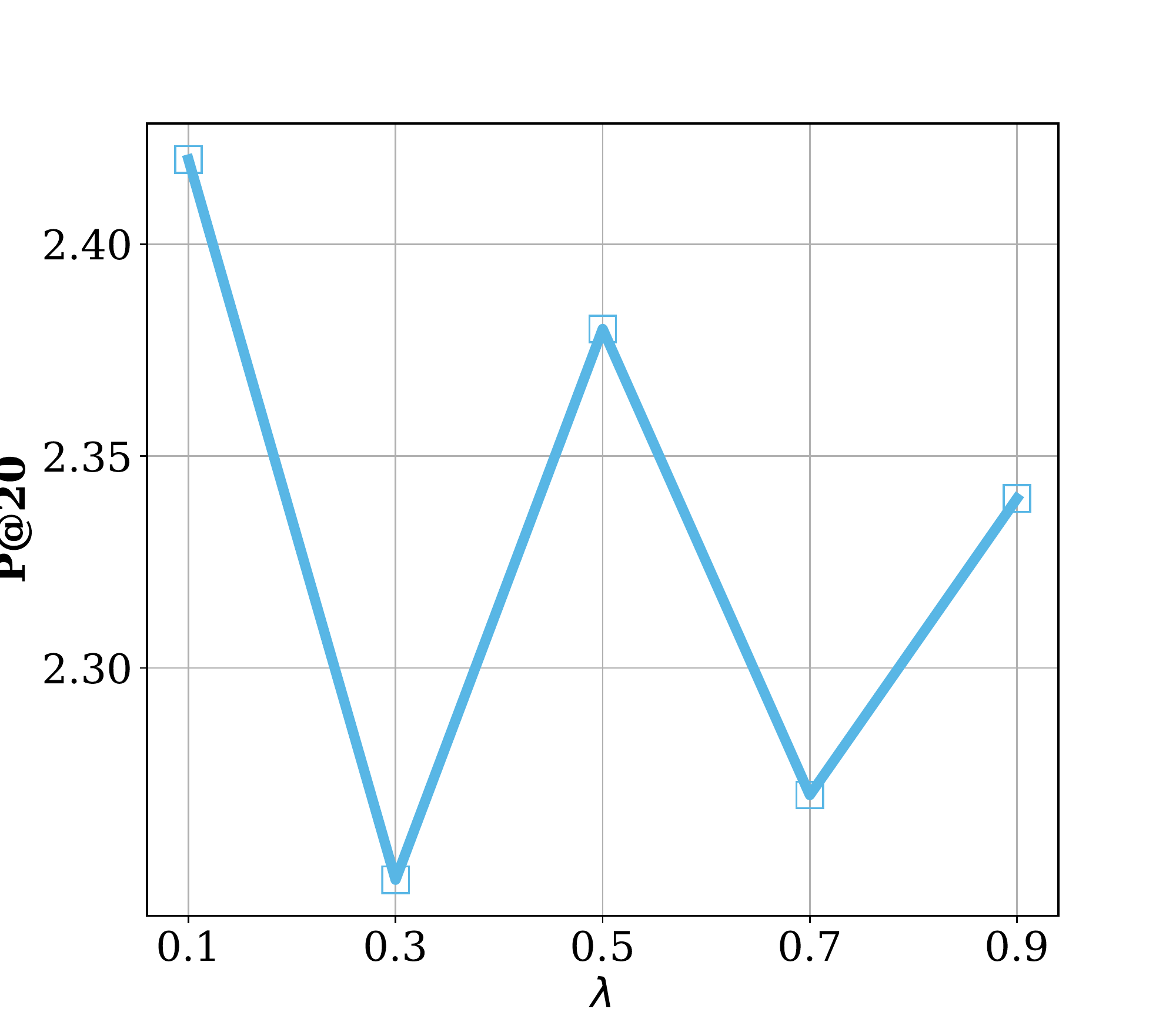}}
        \subfigure[{$\gamma$.}]{\includegraphics[width=0.49\columnwidth]{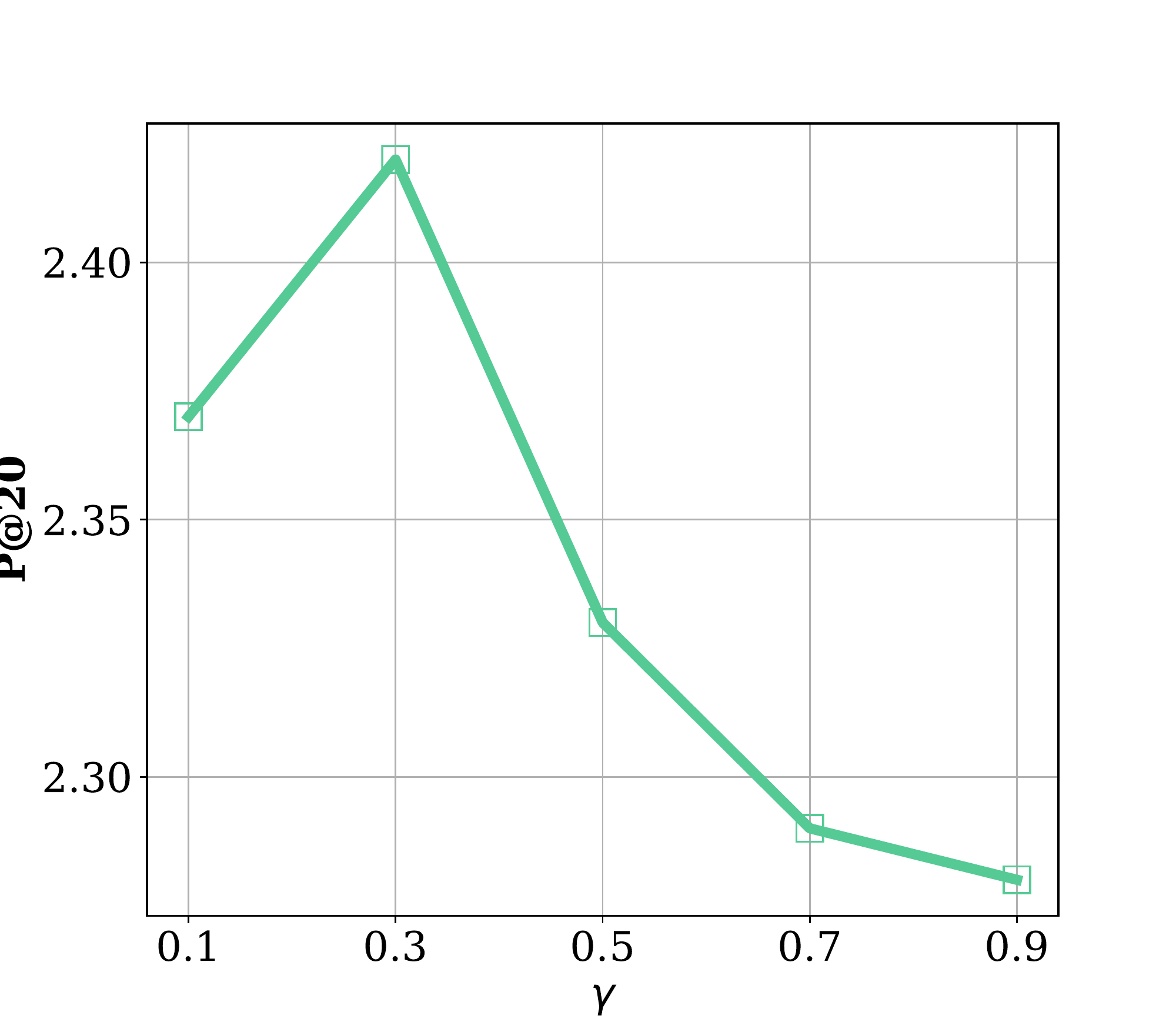}}
	\caption{Parametric sensitivity on $\lambda$ and $\gamma$.}
	\label{lambda}\vspace{-1em}
\end{figure}

\begin{table*}[htpb]
\centering
\caption{Infomation of a session in Meituan dataset.}
\resizebox{1\textwidth}{!}{
\begin{tabular}{@{}l|llllll|lll@{}}
\toprule
\multirow{2}{*}{Item} & \multicolumn{6}{c|}{A historical session}
& \multicolumn{3}{c}{Candidate new items}  \\
 \cmidrule{2-10}
   & 1                & 2                   & 3                    & 4         & 5           & 4          & 6         &\textbf{ 7 }            & 8               \\ 
\midrule
t1         & Gourmet           & Fruit                & Gourment              & Drinks     & Gourmet      & Drinks      & Gourmet    & \textbf{Gourmet}        & Gourmet          \\
t2         & Fast food         & Packaged fruit       & Chinese Cuisine       & Milk Tea   & Fast food    & Milk Tea    & Fast food  & \textbf{Fast food}      & Chinese cuisine  \\
t3         & Hamburger         & Unknown              & Northwestern Cuisine  & Unknown    & Porridge     & Unknown     & Donburi    & \textbf{Meal packages}      & Beijing cuisine  \\
a1         & 1                 & 3                 & 4                 & 2        & 5    & 2         & 6  & \textbf{1 }               & 1  \\ 
a2         & Merchant          & Company              & Company               & Company    & Company      & Company     & Merchant   & \textbf{Company}        & Company  \\

\bottomrule
\end{tabular}
}
\label{meituan}
\end{table*}

\begin{figure*}[t]
	\centering
	\includegraphics[width=0.90\textwidth]{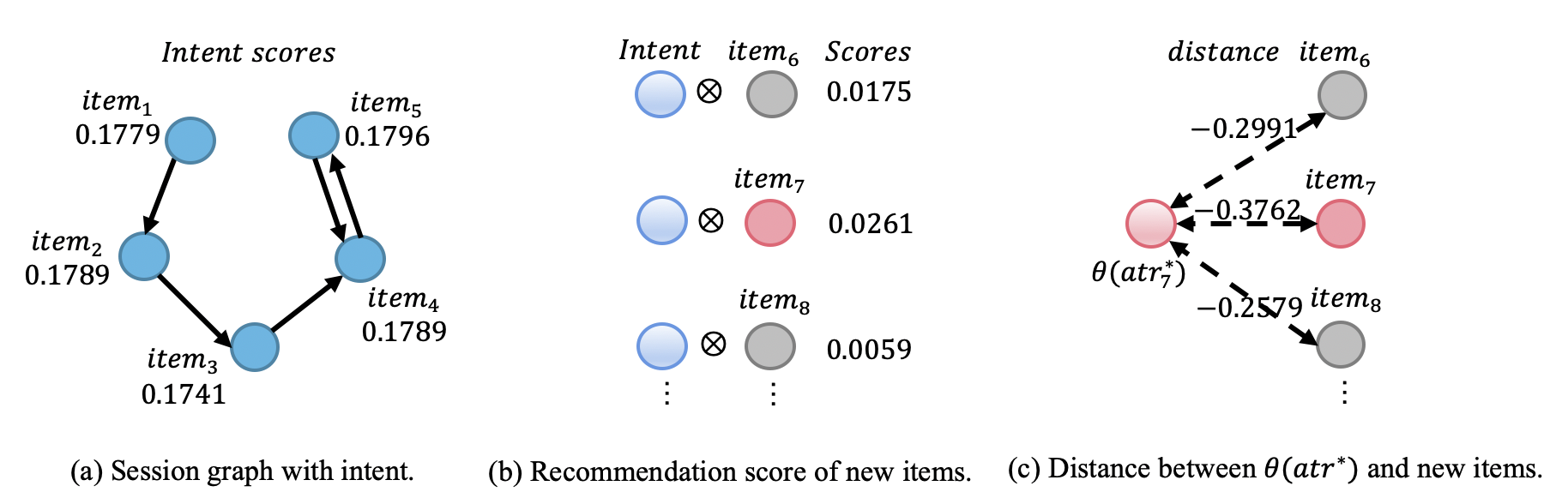} 
	\caption{A commercial case study on the Meituan dataset. }
	\label{mtfig} \vspace{-0.5em}
\end{figure*}

\subsubsection{Ablation Study.}
The ablation study is conducted on the Amazon G\&GF dataset. To clarify the key component of NirGNN, we report the ablation study of the $\alpha$ intent, $\beta$ intent, and the $L_{zero}$ in Table \ref{abs}. \texttt{NirGNN w/o $\alpha$} means that NirGNN is without user $\alpha$ intent and only with user $\beta$ intent. \texttt{NirGNN w/o $\beta$} means that NirGNN is without user $\beta$ intent and only with user $\alpha$ intent. \texttt{NirGNN w/o $L_{zero}$} means that NirGNN is without $L_{zero}$ and only with $L_{ce}$ Loss. The experimental setup is the same as the overall experiment. The results show that our NirGNN model is improved after combining all key components. This reveals the effectiveness of each component.

\subsubsection{Parameter Sensitivity.} Parameter sensitivity analysis experiments are conducted on the Amazon G\&GF dataset. There are two key hyperparameters that should be tuned in our objective function. The first is $\lambda$, which is used to balance $\alpha$ intent and $\beta$ intent in the overall intent $I$. Another hyperparameter is $\gamma$, which is used to balance the cross-entropy loss and $L_{zero}$. By varying $\lambda$ and $\gamma$ from 0.1 to 0.9 in 0.2 intervals, the corresponding hyperparameter sensitivity analysis under $P@20$ metrics is shown in Fig. \ref{lambda}. 
We find that the effect of using a combination of $\alpha$ intent and $\beta$ intent has fluctuated, but is efficacious. The proposed NirGNN model performs best when $\gamma = 0.3$. Although the two intents under the $\lambda$ parameter appear to be conflicting, our ablation experiments demonstrate that in the case of only $\alpha$ or only $\beta$ intent, the effect is not as good as the combination of the two. So they don't cancel each other out.

\subsection{Case Study}
We provide a case study in a real recommendation scenario to demonstrate the interpretability and utility of our model NirGNN. We take a private commercial dataset provided by a food takeaway company, Meituan, as an example to demonstrate the interpretability of the NirGNN model. The dataset is generated by the food orders of Meituan within 30 days in a certain region. The dataset includes a sequence of users buying takeaway food items, a three-level taxonomy tree for each item, and two attributes for each item. We use the dataset provided by Meituan to train the model and randomly select a session from the Meituan dataset to explain our model. Table \ref{meituan} shows information about the case session in the Meituan dataset. The item row shows the ID of each item in this session. $item_{1}$, $item_{2}$, $item_{3}$, $item_{4}$, $item_{5}$ are the takeaway item that the user has visited in the history. $item_{6}$, $item_{7}$, $item_{8}$ are candidate new items, where $item_{7}$ is the ground truth.
$t_1, t_2, t_3$ represent the three-level taxonomy tree, and we show the ID of each taxonomy. $a_1, a_2$ represent the two natural attributes of each takeaway food, including the brand ID and delivery type of the takeaway food. Delivery types include professional delivery (Company) provided by Meituan, and delivery provided by merchants (Merchant).
We provide one session for the case study. This session is $item_{1} \rightarrow item_{2} \rightarrow item{3} \rightarrow item_{4} \rightarrow item_{5} \rightarrow item_{4} $. A user's purchase session is shown in Fig. \ref{mtfig} (a), where $item_{1}$ is the starting item and $item_{4}$ is the ending item. We introduce the interpretability of NirGNN in terms of the interpretability of user intent and the interpretability of reasoning respectively.

\subsubsection{Intent interpretability.} Fig. \ref{mtfig} (a) shows the interpretability of the user dual-intent. In the figure, Each item contains a dual-intent score. We find that $item_{5}$ has the highest score. Intuitively, $item_{4}$ is more in accordance with the user's preferences than others because it is purchased by the user twice recently. However, our intent scores show that the score of $item_{4}$ is lower than $item_{5}$. We observe from Table \ref{meituan} that the data of $item_{5}$ is closer to the ground truth $item_{7}$ than $item_{4}$, which highlights the effectiveness of learning user dual-intent. In addition, Fig. \ref{mtfig} (b) shows the interpretability of the recommendation process. After we learn the user's total intent, the total intent and each candidate new item are used to calculate the matching score for a recommendation. The highest score generated by new items multiplied by user intent will be recommended. We find that the candidate new item $item_{7}$ has the highest score than others under the same intent, so it is considered as the recommended item. The above-mentioned process illustrates the interpretability of the NirGNN recommendation process.

\subsubsection{New item reasoning.} Fig. \ref{mtfig} (c) shows the interpretability of new item embedding reasoning. We put candidate items into the session and obtain their embedding via a GNN encoder. Then we compare the embedding $\theta(atr^*_{7})$ generated by the attributes of $item_{7}$ with the embeddings of other candidate items. We use the Bhattacharyya distance to measure them. We find that the distance between $\theta(atr^*_{7})$ between $item_{7}$ embedding is smaller than the distance between $item_{6}$ or $item_{8}$, which confirms the interpretability of new item reasoning.

\section{Conclusion}
In this paper, we propose a novel GNN-based model, NirGNN, for the GSNIR problem. To effectively learn user intent, we design a dual intent learning network from both the attention perspective and data distribution perspective. At the same time, to solve the problem that the new item has no interaction with the user, we design a zero-shot-based new item reasoning method, which reasons unknown new item embeddings by using attributes of new items. Experimental results show that NirGNN is superior to SOTA methods. A real-world case study shows that NirGNN is interpretable and can be applied in the real world.


\begin{acks} 
This work is supported by the Natural Science Foundation of China under grants 62272340, 62276006 and Meituan Project.
\end{acks}


\bibliographystyle{unsrt}


\end{document}